\documentclass[preprint,showpacs,nofootinbib]{revtex4}
\usepackage{graphicx}

\def\bge{\begin{equation}}
\def\ene{\end{equation}}
\def\bg{\begin{eqnarray}}
\def\en{\end{eqnarray}}

\begin{document}

\title{Response of spin polarized neutron matter under the presence of a strong magnetic field with Skyrme interactions.}

\author{M. \'{A}ngeles P\'{e}rez-Garc\'{i}a~\footnote{mperezga@usal.es}}

\affiliation{$^{1}$Departamento de F\'{i}sica Fundamental and IUFFyM,\\ Universidad de Salamanca, Plaza de la Merced s/n 37008 Salamanca}

\date{\today}

\begin{abstract}
The effects of a strong magnetic field in the dynamical response of a pure neutron system to a weak neutrino probe are studied within the framework of the Landau Fermi Liquid Theory in the non-relativistic Hartree-Fock  approximation. We use Skyrme forces to parametrize the partially magnetized nuclear interacting plasma and describe its collective modes. We study the vector and vector-axial matter response functions and the neutrino mean free path in this two-component fermion system with net spin polarization and density dependent in-medium correlations. We find a decrease in the neutrino opacity of magnetized matter with respect to the non-magnetized case for fields $log_{10}\, B(G) \gtrsim 17$. 
\vspace{1pc}
\end{abstract}

\maketitle

\section{Introduction}
\label{intro}

The theoretical study of properties of hadronic matter under extreme conditions allows to understand nuclear systems ranging from experimental heavy ion collisions \cite{heavyion} to the astrophysical scenarios with objects such as neutron stars, where densities up to several times that of nuclear saturation are believed to exist in their inner core \cite{neutronstar}. The study of the low temperature behaviour of Fermi Liquids was pioneeringly addressed by Landau \cite{landau} and later developed by others in the Fermi Liquid Theory (FLT) \cite{bookFL}. The main idea underlaying is that the properties of this type of normal quantum systems can be studied using the dynamics of quasiparticles close to the Fermi surface. 
The inclusion of an additional magnetic field, $B$, allows further testing the properties of magnetized Fermi Liquids since nucleons have a tiny, but not vanishing, magnetic moment. 

These degenerate plasmas are of interest in the description of matter in the inner shells of rapidly rotating neutron stars with very intense magnetic fields. In the catalog of the $\approx 1,500$ such objects there is now experimental indication that in a subgroup of pulsars, called {\it magnetars}, magnetic field strengths in the surface of the order $B \approx 10^{15}$ G \cite{thom, lazzati} exist. In the formation of these objects, after a Supernova core collapse, neutrinos are an important ingredient in the first stages of the cooling of the proto-neutron star \cite{cooling}. The interactions of (anti) neutrinos with matter determine the ratio of deposited energy that, in turn, affect the dynamics of the cooling afterwards.  

As a first approach, and in most of the calculations, the modifications due to the induced spin polarization in nuclear systems under the presence of magnetic fields are discarded. For example, some works have calculated the matter response functions in absence of magnetic fields, first, for the pure neutron system \cite{pnm1, pnm2, pnm3} or in the case where some proton population is included \cite{npa-nav1,npa-nav2,annals, works_skyrme1,works_skyrme2,works_skyrme3,works_skyrme4,works_skyrme5}. For systems where a spin excess is postulated {\it a priori} some partial results have been obtained in the context of Hartree-Fock calculations as in \cite{vida1}. However, in these works there is no link between a magnetic field and the induced matter  magnetization.

 It has been argued that it is important to know whether a ferromagnetic transition takes place or not  \cite{jerome} in nuclear systems at high densities and the impact of the spin polarization as a source of additional excitation modes in the magnetized plasma that, in turn, could affect the neutrino opacity in the nuclear medium. To describe the nuclear interaction, a diversity of  models in the relativistic \cite{prakash} or  phenomenological non-relativistic \cite{works_skyrme1} approximations have been considered. On the other hand, in the low density case, recent works show that the vector response seems to be relevant in the low density non-magnetized matter and  direct simulations either using Monte Carlo or Molecular Dynamics techniques have been performed. At these low densities clustered pasta phases may develop~\cite{pasta1,pasta2,pasta3,pasta4}.

In an earlier work \cite{angprc} we showed that there are important modifications in the low and high density regimes, as magnetization of a neutron plasma is concerned, when one considers phenomenological nuclear interaction models, such as Skyrme and Gogny forces. At supranuclear densities, $1-3$ times saturation density, the residual magnetization amounts to values typically below $5\%$ for magnetic fields as high as $B \approx 10^{18}$ G allowed by the scalar virial theorem and moderate temperatures. We have also calculated the influence of the generalized Landau Fermi liquid parameters \cite{notes} on a set of observables in a magnetized neutron system \cite{ang2}.

The present work performs a study of dynamical response functions for a magnetized neutron plasma and the effect of a strong magnetic field on the neutrino opacities in the two-component neutron Fermi Liquid, using Skyrme effective nuclear interactions. 
The structure of the paper is as follows. In section~\ref{model}, we describe the medium response under the linear approximation for the different parametrizations of the Skyrme force used in the description of the nucleon-nucleon (NN) interaction. Using the Landau Fermi Liquid Theory at zero temperature we consider a multipolar expansion of the particle-hole (ph) interaction and analyze the linear response up to dipolar terms.  We gauge the effects of finite density, presence of a strong external magnetic field and the induced spin polarization in the neutron system. In section~\ref{results} we present the results obtained and finally summary and conclusions are given in section~\ref{summary}.

\section{Response functions in the magnetized neutron Fermi Liquid}
\label{model}

We consider a homogeneous pure neutron system where each neutron has spin projection on the  z-axis, $\sigma$. It can be either $\sigma=+1$ for spins aligned parallel to a uniform magnetic field that will be taken along the z-direction, ${\bf B}=B {\bf k}$, or $\sigma=-1$ for antiparallel spins. This system may be viewed as a two-component neutron Fermi Liquid where the relative populations of magnetized species vary as magnetic field and density change. Although we restrict ourselves to neutron matter in this work, for application to astrophysical scenarios, i. e., in neutron star matter \cite{nstar} beta equilibrium  must be imposed and, subsequently, multi-component plasmas with leptons, protons and heavier baryons should be considered.

The total baryonic density in our two-component neutron system is given by $\rho=\rho_+ + \rho_-$, being $\rho_{\sigma}=\frac{k^3_F(\sigma)}{6 \pi^2}$ the $\sigma$-polarized component at $T=0$. In this kind of systems there may be some spin excess, $\Delta$, defined as $\Delta=\frac{\rho_+ - \rho_-}{\rho}$.
As explained in \cite{angprc} for given conditions of density, $\rho$, temperature (we fix here $T=0$) and a strong magnetic field strength, $B$, the relevant thermodynamical potential is the Helmholtz free energy, $F_M$, defined as~\cite{callen}  
\begin{equation}
F_M=E-MB,
\label{Fm}
\end{equation}
where E is the energy and $M=\int m dV$ is the total magnetization of a given volume. The net magnetization density is $m=\mu_n \Delta \rho$ and $\mu_n=-1.9130427(5)\mu_N$ is the neutron magnetic moment in units of the nuclear magneton~\cite{pdb}. Note that parallel (antiparallel) aligned magnetic moments (spins) are energetically favoured.

In this work we have considered an effective approach to the nuclear interaction using two different parametrizations of the phenomenological zero-range Skyrme force. In the usual way it can be written as~\cite{vautherin} 
\begin{eqnarray}
V^{Skyrme}_{NN}({\bf r}_1,{\bf r}_2)&=& t_0 \left(1+x_0 P^{\sigma} \right) \delta({\bf r})  + \frac{1}{2} t_1 \left(1+x_1 P^{\sigma} \right)
\left[ {\bf k'}^2 \delta({\bf r}) +  \delta({\bf r}) {\bf k}^2 \right] 
\nonumber \\
&& +  t_2 \left(1+x_2 P^{\sigma} \right) {\bf k'} \cdot \delta({\bf r}) {\bf k}
+ \frac{1}{6} t_3 \left(1+x_3 P^{\sigma} \right) \rho^{\alpha} ({\bf R})\delta({\bf r}) 
\label{skyrme}
\end{eqnarray}
where ${\bf r}={\bf r}_1-{\bf r}_2$ and ${\bf R}=({\bf r}_1+{\bf r}_2)/2$, 
${\bf k}=({\bf \nabla}_1-{\bf \nabla}_2)/2i$ the relative momentum acting on the right and ${\bf k'}$ its conjugate acting on the left. $P^{\sigma}$ is the spin exchange operator. Note that we have omitted terms not relevant for homogeneous systems.
From the large myriad of Skyrme parametrizations we have chosen the widely used SLy4 and SLy7, given by the Lyon group~\cite{chabanat1,chabanat2} as they provide good values for binding of nuclei and also for  neutron matter equation of state (EOS) giving values of maximum neutron star masses around $1.5 M_{\odot}$~\cite{skyrme2, skyrmeeos}. Saturation density is slightly different for both parametrizations, being  for the SLy4 (SLy7) case $0.158\, fm^{-3}$  ($0.160\, fm^{-3}$ ).

In the context of the FLT the properties of these systems at low temperature are related to the dynamics of quasiparticle excitations with momentum $k$ and spin projection, $\sigma$,  around the Fermi surface in the presence of a magnetic field.  These are studied by evaluating the quasiparticle interaction matrix element~\cite{bookFL} and will be a crucial ingredient needed in order to compute the response of the pure and possibly magnetized hadronic system. They can be obtained for a non-magnetized system \cite{plbbackman} from the usual multipolar expansion  in Legendre polinomials of the quasiparticle interaction and we can write,
\begin{equation}
V_{ph}=\sum_{l=0}^{\infty} \big [ f_l + g_l {\bf \sigma_1 .\sigma_2} \big ] P_l ( cos\theta) ,
\label{elemento}
\end{equation}
where $\theta$ is the angle related to the interacting quasiparticle three-momenta  and $f_l$ and $g_l$ are the so-called Landau parameters. In the more general case where two possible spin orientations $\sigma=\pm 1$ are taken into account, the polarized quasiparticle matrix elements can be written \cite{notes} using coefficients $f_l^{(\sigma,\sigma')}$ depending on the degree of polarization, $\Delta$, and on the Fermi momenta $k_F^{\sigma}$, $k_F^{\sigma'}$ involved. In the limit $\Delta \rightarrow 0$ the Landau parameters fullfill the following relations,
\begin{equation}
f_l=\frac{f_l^{(\sigma,\sigma)}+f_l^{(\sigma,-\sigma)}}{2},
\label{f0sum}
\end{equation}
\begin{equation}
g_l=\frac{f_l^{(\sigma,\sigma)}-f_l^{(\sigma,-\sigma)}}{2}.
\label{g0sum}
\end{equation}
For the effective Skyrme interaction that we will be considering in this work the only non-vanishing  terms are the monopolar ($l=0$) and the dipolar ($l=1$) ones. In the Skyrme model they are independent of temperature, so that they just retain the density and polarization dependence. They can be written for the monopolar case as ($\sigma'=\pm \sigma$),
\begin{eqnarray}
f_0^{(\sigma,\sigma)} &=& \frac{1}{6} t_3 (1-x_3) (\alpha (\alpha-1) \rho^{\alpha-2} \rho_+ \rho_+ + 2 \alpha \rho^{\alpha-1}( \rho - \rho_{\sigma} ))-  \nonumber \\ && + t_2 (1+x_2) k^2_{F,\sigma} ,
\end{eqnarray} 
\begin{eqnarray} 
f_0^{(\sigma,-\sigma)} &=& t_0 (1-x_0) + \frac{1}{6} t_3 (1-x_3)\big [ \alpha (\alpha-1) \rho^{\alpha-2} \rho_+ \rho_- + (\alpha+1) \rho^{\alpha}\big ] -  \nonumber \\ && + \frac{1}{4} (t_1 (1-x_1) + t_2 (1+x_2) )\left( k^2_{F,\sigma} + k^2_{F,-\sigma} \right) ,
\end{eqnarray}
while the dipolar terms are given by, 
\begin{equation} 
f_1^{(\sigma,\sigma)} = -t_2 (1+x_2) k^2_{F,\sigma} ,
\end{equation} 
\begin{equation} 
f_1^{(\sigma,-\sigma)} = -\frac{1}{2} (t_1 (1-x_1) + t_2 (1+x_2) ) k_{F,\sigma}\rm k_{F,-\sigma}. 
\end{equation}  
In the Skyrme interaction the quasiparticle effective mass in a magnetized system depends on the polarized dipolar coefficients,
\begin{equation} 
 m^*_{\sigma}/m=1+\frac{1}{3} N_{0 \sigma} \big [ f_1^{(\sigma,\sigma)}+(\frac{k^2_{F,-\sigma}} {k^2_{F,\sigma}})f_1^{(\sigma,-\sigma)} \big ] 
\label{mef}
\end{equation}
where  $N_{0 \sigma}=\frac{m^{*}_{\sigma} k_{F,\sigma}}{2 \pi^2}$ contains also $m_{\sigma}^*$ and are the quasiparticle  level densities at each polarized Fermi surface. The response functions in a neutron system for a weakly interacting neutrino probe under the presence of a magnetic field $B$ can be obtained from the Landau FLT  through the matter susceptibilities for a given multipolarity, $l$ \cite{rpa}. Accordingly, the Lindhard function, $\chi^{(\sigma ,\sigma')}=\chi^{(\sigma ,\sigma')}(\omega, q)$, satisfy the Bethe-Salpeter equation and can be written under the form of an algebraic system that, in the monopolar random phase approximation (RPA), reads as a $4 \times 4$ matrix,
\begin{equation}
\chi^{(\sigma ,\sigma')} = \chi_0^{(\sigma)} \delta(\sigma, \sigma') +  
\chi_0^{(\sigma)} \sum_{\sigma''=+,-} f_0^{(\sigma \sigma'')}
\chi^{(\sigma'' \sigma')},
\label{chil0}
\end{equation}

where $\chi^{\sigma}_0=\int \frac{d^3 k}{(2 \pi)^3}G_0^{\sigma}$ corresponds to the non-correlated response due to the $\sigma$-polarized neutron fraction obtained from integration of the quasiparticle propagator $G_0^{\sigma}$. In the limit where the energy transfer is close to zero, $\omega \rightarrow 0$, and $\omega/q$ is approximately constant, then $Re[\chi^{\sigma}_0] \rightarrow -N_{0 \sigma}$.

The explicit solution in the RPA monopolar $(l=0)$ case can be written as,
\begin{eqnarray}
\chi^{(+, \sigma')} &=& \frac{1}{D} \left(
\chi_0^{(+)} \delta(\sigma',+) - \chi_0^{(+)} \chi_0^{(-)} \left[
f_0^{(--)} \delta(\sigma',+) - f_0^{(+-)} \delta(\sigma',-) \right] \right), \\
\chi^{(-, \sigma')} &=& \frac{1}{D} \left(
\chi_0^{(-)} \delta(\sigma',-) - \chi_0^{(+)} \chi_0^{(-)} \left[
f_0^{(++)} \delta(\sigma',-) - f_0^{(-+)} \delta(\sigma',+) \right] \right) ,
\end{eqnarray} 
where the determinant of the set is
\begin{equation}
D = 1 - \chi_0^{(+)} f_0^{(++)} - \chi_0^{(-)} f_0^{(--)}
+ \chi_0^{(+)} \chi_0^{(-)} \left( f_0^{(++)} f_0^{(--)} - 
f_0^{(+-)} f_0^{(-+)} \right).
\end{equation}
In the dipolar approximation ($l \le 1$) we can write the RPA algebraic coupled system including the susceptibility as a $8 \times 8$ matrix \cite{notes}
\begin{equation}
\chi^{(\sigma ,\sigma')} = \chi_0^{(\sigma)} \delta(\sigma, \sigma') +  
\chi_0^{(\sigma)} \sum_{\sigma''=+,-} f_0^{(\sigma \sigma'')}
\chi^{(\sigma'' \sigma')}+ \gamma_1^{(\sigma)} \sum_{\sigma''=+,-} f_1^{(\sigma, \sigma'')}
\Gamma^{(\sigma'', \sigma')},
\label{chil1}
\end{equation}
\begin{equation}
\Gamma^{(\sigma ,\sigma')} = \gamma_1^{(\sigma)} \delta(\sigma, \sigma') +  
\gamma_1^{(\sigma)} \sum_{\sigma''=+,-} f_0^{(\sigma \sigma'')}
\chi^{(\sigma'' \sigma')}+ \gamma_2^{(\sigma)} \sum_{\sigma''=+,-} f_1^{(\sigma, \sigma'')}
\Gamma^{(\sigma'' \sigma')},
\label{chig2}
\end{equation}
where we have defined the following auxiliar quantities
\begin{equation}
\Gamma^{(\sigma, \sigma')}=\int \frac{d^3 k}{(2 \pi)^3} cos (\theta)\, G^{(\sigma,\sigma')},
\label{gamma}
\end{equation}
\begin{equation}
\gamma_n^{(\sigma)}=\int \frac{d^3 k}{(2 \pi)^3} cos^n (\theta)\, G_{0}^{(\sigma)}.
\label{gamma_n}
\end{equation}
The expressions for the coefficients $\gamma_i^{(\sigma)}$ can be written in the Landau limit as
\begin{equation}
\gamma_1^{(\sigma)}= \nu^{(\sigma)} \chi^{(\sigma)}_0,
\label{gamma1}
\end{equation}
\begin{equation}
\gamma_2^{(\sigma)}=\nu^{2 (\sigma)} \chi^{(\sigma)}_0-\frac{k_{F,\sigma} m^{*}_{\sigma}}{6 \pi^2},
\label{gamma2}
\end{equation}

where $\nu^{(\sigma)}=\frac{m^{*}_{\sigma} \omega}{k_{F,\sigma} q }$.  
In-medium effects are incorporated by the polarized quasiparticle propagator, $G^{(\sigma)}_0$, and corrections to that at the linear approximation through $G^{(\sigma,\sigma')}$, obtained by summing all the ring diagrams corresponding to the magnetized ph excitations calculated at a given $l$-multipolarity.
Then, the Lindhard function is calculated from the propagator  by integrating over the incoming neutrino momentum as,
\begin{equation}
\chi^{(\sigma,\sigma')}(\omega,q)= \int \frac{d^3 k}{(2 \pi)^3} G^{(\sigma,\sigma')}.
\label{chi1}
\end{equation}

Having obtained the generic reponse in the $(\sigma,\sigma')$ channel, we can now constrain to the isovector contribution, $S=0$, as,
\begin{equation}
\chi^{(S=0)} =  \chi^{(++)} + \chi^{(--)} + \chi^{(+-)} + \chi^{(-+)} ,
\end{equation}
and the vector axial contribution, $S=1$, reads,
\begin{equation}
\chi^{(S=1)} = \chi^{(++)} + \chi^{(--)} - \chi^{(+-)} - \chi^{(-+)}.
\end{equation}
The structure functions can be calculated from the Lindhard function  at zero temperature as \cite{rpa},
\begin{equation}
S^{S=0,1}(\omega,q)=\frac{-1}{\pi} Im \,\chi^{S=0,1}(\omega,q).
\label{Imchi}
\end{equation}
In this case of vanishing temperature only non-negative energy transfers may happen in the polarized neutron system. Then, the response function $S^{(S)}$ will provide information about the density-dependent in-medium correlations, partial polarization of the plasma, and vector and vector-axial contributions to the overall response of matter that may arise in the presence of a strong magnetic field when it is weakly probed by neutrinos. For example, in the neutral current reaction involving $\nu_{\mu},\nu_{\tau}$ neutrinos, collectively $\nu_x$, and their antiparticles, elastically scattering  up (down) polarized neutrons,
\begin{equation}
\nu_x + n^{\sigma} \rightarrow \nu_x + n^{\sigma}.
\end{equation}
In the non-relativistic limit the differential cross section of neutrinos scattering off matter can be calculated from the linear response of the medium using \cite{sawyer}
\begin{equation}
\frac{1}{V} \frac{d^3 \sigma} {d \Omega d \omega}=\frac{G^2_F}{8 \pi^3} E'^2 (1-f_{\nu}(E')) 
\{ C_V^2 (1+cos \theta) S^{0}(\omega,q)+C_A^2 (3-cos \theta)S^{1}(\omega,q)\}.
\end{equation}
where $E$ and $E'$ are the incoming and outgoing (anti) neutrino energies, respectively. The transferred energy is $\omega=E-E'$ and the transferred three-momentum is obtained from the (anti) neutrino incoming   $(\vec{k})$ and outgoing  $(\vec{k'})$ three-momentum as $\vec{q}=\vec{k} -\vec{k'}$. The neutral current vector and axial vector charges are $C_V=1/2$ and $C_A=-g_a/2$ where $g_a=1.260$ \cite{pdb}. $G_F/(\hbar c)^3=1.166\,39(1) \times 10^{-5} GeV^{-2}$ is the Fermi coupling constant. 
Once the response has been evaluated it is straightforward to evaluate the neutrino mean free paths in the medium using 
\begin{equation}
\lambda^{-1}=\int \frac{1}{V} \frac{d \sigma} {d \Omega d\omega} d \Omega d \omega.
\end{equation}
Let us mention that, although the low temperature case in an astrophysical scenario arises after the early cooling neutrino phase, it is worth sizing the dependence of the neutrino opacity on medium effects, spin polarization and strong magnetic fields with respect to the non-magnetized case in an analogous way to the study of the ideal cases of isospin symmetric nuclear matter and pure neutron matter evaluated before addressing the more realistic situation where different hadronic species in beta equilibrated matter arise.

\section{Results}
\label{results}

\begin{figure}[hbtp]
\begin{center}
\includegraphics [scale=1.5] {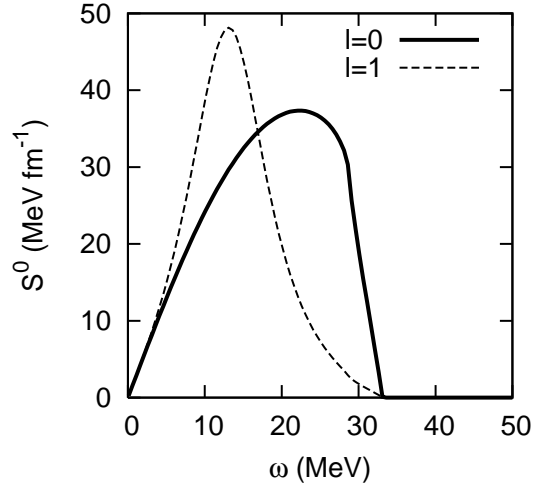}
\caption{Vector response function $S^0$ at $q=0.25$ $fm^{-1}$ for a magnetic field $B=10^{17}$ G at saturation density with the Skyrme SLy7 foce for $l=0$ and $l=1$.} 
\label{Fig1}
\end{center}
\end{figure}

\begin{figure}[hbtp]
\begin{center}
\includegraphics [scale=1.5] {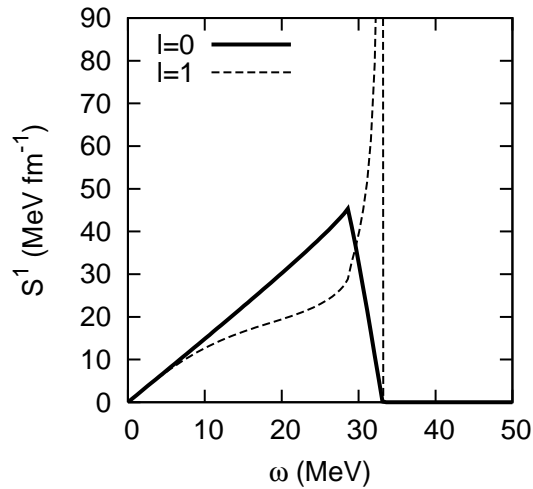}
\caption{ Spin response function $S^1$ for the same conditions as in Fig.~\ref{Fig1}.} 
\label{Fig2}
\end{center}
\end{figure}

\begin{figure}[hbtp]
\begin{center}
\includegraphics [scale=1.5] {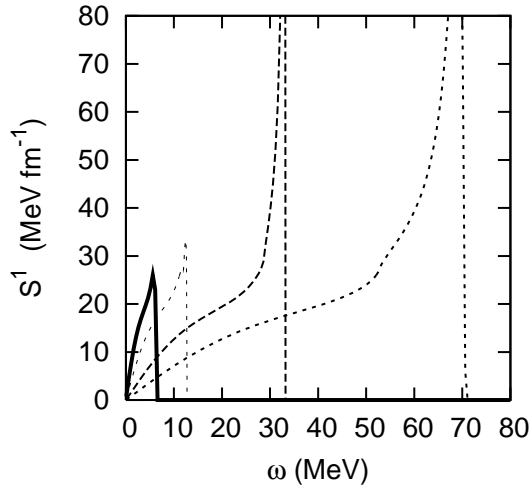}
\caption{Spin response function $S^1$ in the RPA dipolar approximation for values (from left to right) of $q=0.05,0.1,0.25,0.5$ $fm^{-1}$ for a magnetic field $B=10^{17}$ G at saturation density with the Skyrme SLy7 parametrization.} 
\label{Fig3}
\end{center}
\end{figure}

In this section we present results arising from a consistent treatment \cite{angprc} minimizing the Helmholtz free energy  calculated in Hartree-Fock approximation under the presence of a strong magnetic field $F_M(\Delta, \rho, B)$. We have considered magnetic fields up to a maximum {\it internal} value  $B_{max} \approx 10^{18}$ G according to the scalar virial theorem. 
In Fig.~\ref{Fig1} and Fig.~\ref{Fig2} we plot the  vector, $S^{0}(\omega, q)$, and spin, $S^{1}(\omega, q)$, response functions respectively, as a function of the energy transfer, $\omega$, for the Skyrme SLy7 interaction at saturation density $\rho_0$ in the RPA monopolar (solid line) and RPA dipolar (dashed line) approximations at a value of the transferred three momentum, $q=0.25\, fm^{-1}$. We can see that dipolar terms add non-negligible contributions to the monopolar response functions due to the fact that they retain the energy and momentum dependence of the residual interaction. In the spin density channel the effect is a shift in the collective mode to lower energies while in the vector channel is the opposite. 
In Fig.~\ref{Fig2} the spin response function shows the presence of spin zero sound modes as obtained in the dipolar approximation. Under these conditions there is a very mild polarization ($\Delta \approx -1\%$) of the neutron plasma. The collective spin zero sound modes are very close and their shift is hardly visible on the plot, $\omega_+ \approx \omega_- \approx 33.0$ MeV. The ph limit for the $\sigma$-polarized population is given by  $\omega_{\sigma}=\frac{q(q+2 k_{F,\sigma})}{2 m^{*}_{\sigma}}$. 
In Fig.~\ref{Fig3} we can see the  spin response function, $S^{1}$, in the RPA dipolar approximation as a function of the energy transfer, $\omega$, for the Skyrme SLy7 interaction at saturation density, $\rho_0$, and zero temperature for different values of the transferred three momentum, $q=0.05\, fm^{-1}$ (solid line), $q=0.1\, fm^{-1}$ (short dashed line), $q=0.25\, fm^{-1}$ (long dashed line) and $q=0.5\, fm^{-1}$ (dotted line) respectively. For this plot we have set the magnetic field strength $B=10^{17}$ G. We can see that the ph continuum is achieved at a value $max[\omega_+,\omega_-]$ and again at this density the very mild polarization of the system results in the relative separation of the modes to be less than $1\%$ although increasingly larger as $q$ increases.

\begin{figure}
\begin{minipage}[b]{0.5\linewidth} 
\centering
\includegraphics[scale=1.]{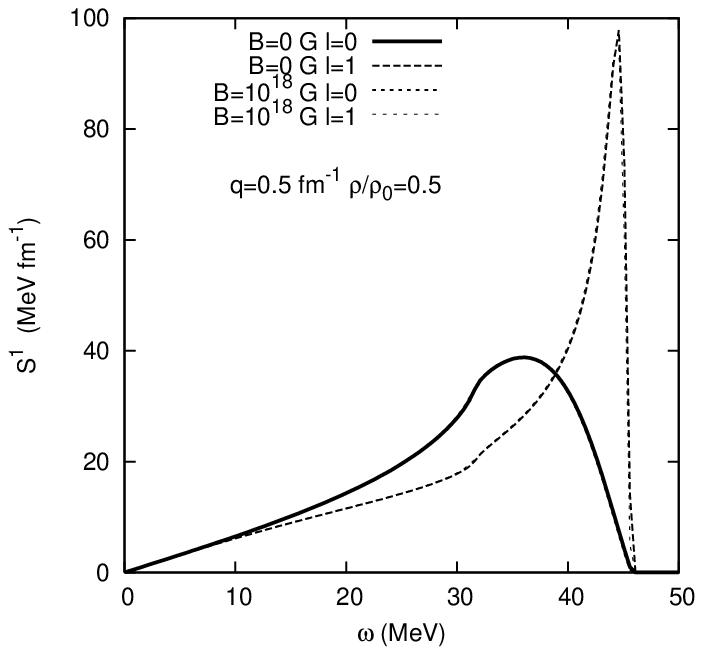}
\end{minipage}
\hspace{0.5cm} 
\begin{minipage}[b]{0.5\linewidth}
\centering
\includegraphics[scale=1.]{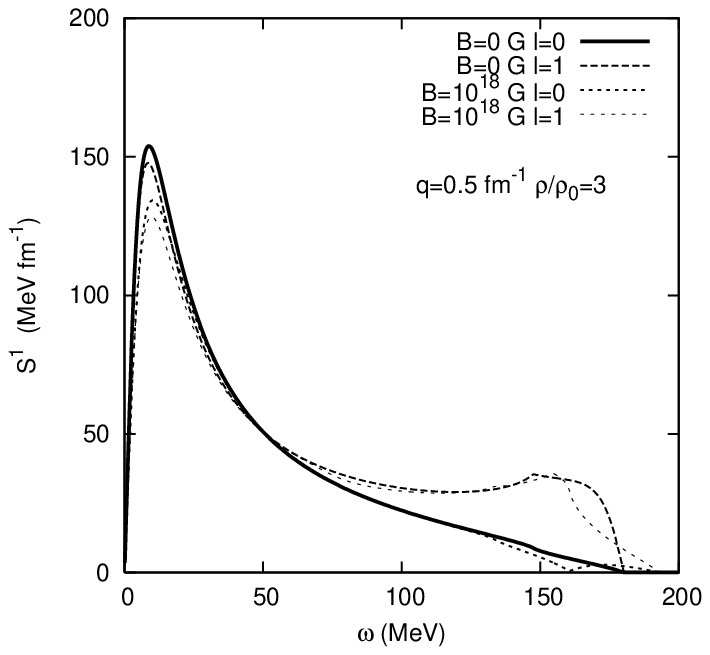}
\caption{Spin response function at densities $\rho=0.5\rho_0$ (upper panel) and $\rho=3 \rho_0$ (lower panel) with the SLy7 model in the RPA $l=0,1$ cases for limiting values of $B$. See text for details.} 
\label{Fig4}
\end{minipage}
\end{figure}

In Fig.~\ref{Fig4} we can see the spin response function $S^{1}$ as a function of the energy transfer, $\omega$, for the Skyrme SLy7 interaction at $\rho=0.5 \rho_0$ (upper pannel) and $\rho=3\rho_0$ (lower pannel) in the RPA monopolar $(l=0)$ and RPA dipolar $(l=1)$ approximations for a value of the transferred three momentum, $q=0.5\, fm^{-1}$. In order to size the effect of a magnetic field we have studied limiting cases ranging from $B=0$ G to the maximum field strength considered in this work, $B=10^{18}$ G. We can see that in the low density case there is a high energy mode shifted in the dipolar case with respect to the monopolar case. However the inclusion of magnetic field produces an undistiguishable (on the plot) shift in the strength due to the low induced polarization at this density \cite{angprc}. However, at higher density a low energy mode arises, since a ferromagnetic transition is near. The presence of magnetic field tends to decrease (increase) the low (high) energy strength of the response slightly. At low densities a realistic treatment would include the possible existence of magnetic pasta phases analogous to the clustered configurations obtained in \cite{pasta1, pasta2, pasta3,pasta4}.

\begin{figure}[hbtp]
\begin{center}
\includegraphics [scale=1.5] {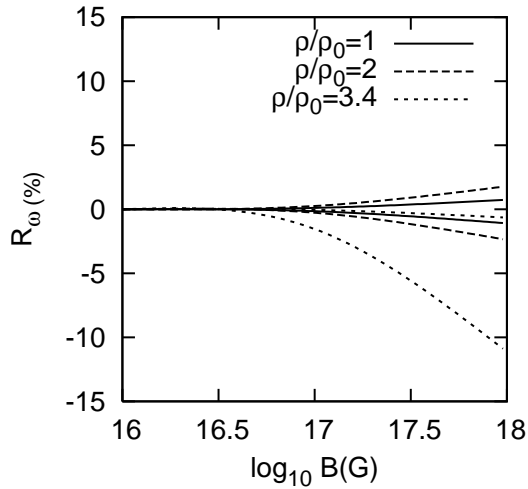}
\caption{Ratio of change of the ph continuum energy (in percentage) versus the logarithm of the magnetic field strength in the RPA dipolar approximation at different densities, $\rho/\rho_0=1, 2, 3.4$ and $q=0.25\, fm^{-1}$ with the Skyrme SLy7 interaction. See text for details.} 
\label{Fig5}
\end{center}
\end{figure}

\begin{figure}[hbtp]
\begin{center}
\includegraphics [scale=1.5] {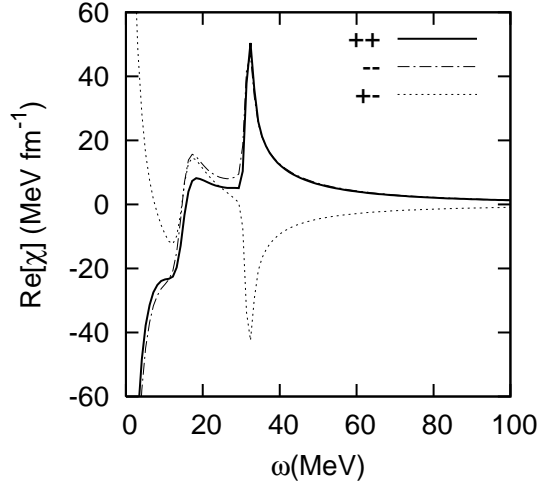}
\caption{Real part of the Lindhard function versus energy transfer as calculated with the Skyrme SLy4 interaction with $B=5 \times 10^{17}$ G in the RPA dipolar approximation, $\rho/\rho_0=3$ and $q=0.1\, fm^{-1}$.} 
\label{Fig6}
\end{center}
\end{figure}

\begin{figure}[hbtp]
\begin{center}
\includegraphics [scale=1.5] {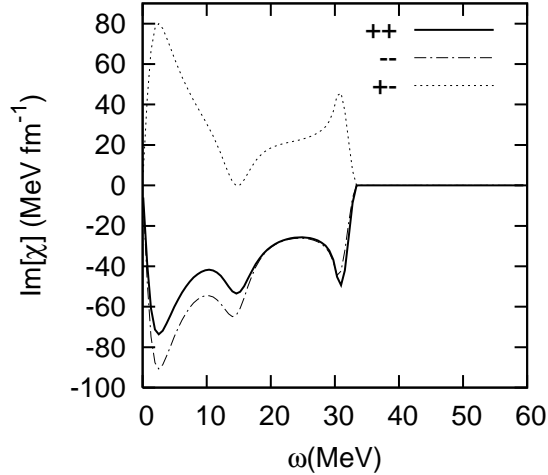}
\caption{Same as Fig.~\ref{Fig6} but for the imaginary part.} 
\label{Fig7}
\end{center}
\end{figure}

In order to further clarify the contribution of the magnetic field to the change in the spin response function we plot on Fig.~\ref{Fig5} the ratio of change of the ph continuum for the $\sigma$-polarized population component with respecto to the $B=0$ G value, $\omega_0$ at a given value of $q$, that we define as $R^{\sigma}_{\omega}=\frac{{\omega}_{\sigma}-\omega_0}{\omega_0}$ versus the logarithm (base 10) of the magnetic field strength for densities $\rho/\rho_0=1, 2, 3.4$ in solid, long dashed and short dashed line respectively and $q=0.25\, fm^{-1}$ in the RPA dipolar approximation. The change for fields below $B \approx 10^{16}$ G is almost negligible. We can see on the plot that the ratio is positive (negative) for $R^{\sigma}_{\omega}$ for the parallel (antiparallel) spin populations in the $\rho/\rho_0=1, 2$ cases. In the $\rho/\rho_0=3.4$ and due to the proximity of the ferromagnetic phase transition both ratios are negative and the antiparallel collective mode is largely affected to  about $\approx -11 \%$. 
In Fig.~\ref{Fig6} and Fig.~\ref{Fig7} we show the real and imaginary parts, respectively, of the dynamical  Lindhard function, $\chi^{(\sigma ,\sigma')}$, as calculated with the Skryme SLy4 interaction at $\rho=3 \rho_0$ with the RPA dipolar approximation for a value of $B=5 \times 10^{17}$ G and $q=0.1\, fm^{-1}$. We can see that in the real and imaginary parts there is a shift to the lower energies in the dominant spin down component ($\approx 10\%$) and low energy modes are excited since at this density a ferromagnetic transition is near. For the imaginary part both the $(+,+)$ and $(-,-)$ component are negative while the mixing part $(+,-)$ gives positive contribution so that the structure function in the $S=1$ (S=0) channel is increased (decreased). As density grows   the $\omega_+$ and $\omega_-$ modes drift away from each other since the spin polarization of the plasma is larger. 

\begin{figure}[hbtp]
\begin{center}
\includegraphics [scale=1.5] {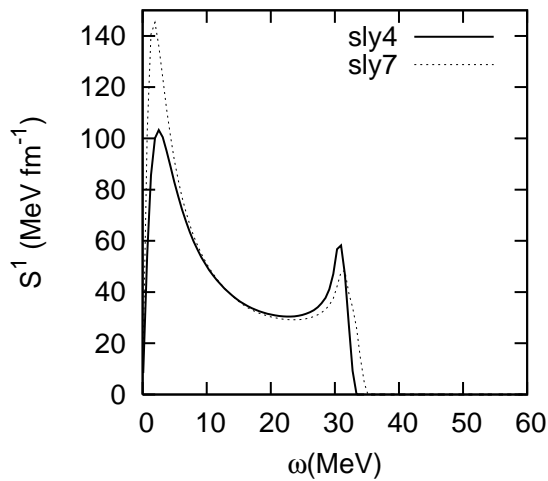}
\caption{Spin response function versus energy transfer for the Skyrme SLy4 and SLy7 parametrizations at $B=5 \times 10^{17}$ G in the RPA dipolar approximation, $\rho/\rho_0=3$ and $q=0.1\, fm^{-1}$.} 
\label{Fig8}
\end{center}
\end{figure}

\begin{figure}[hbtp]
\begin{center}
\includegraphics [angle=-90,scale=.75]{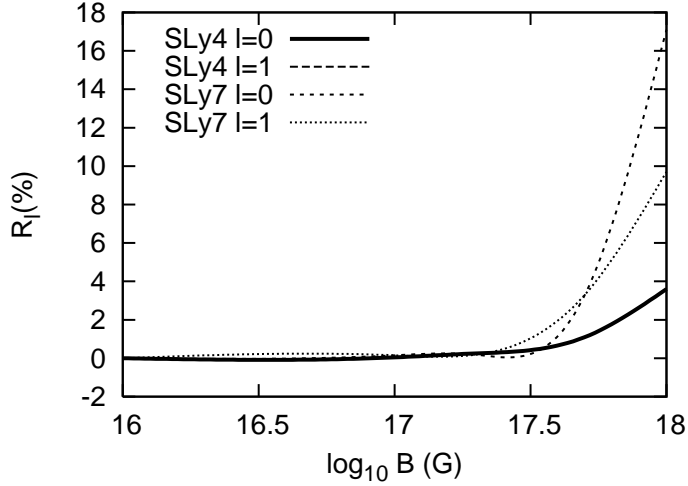}
\caption{Ratio of neutrino mean free paths calculated with the SLy4 and SLy7 in the RPA approximation with $l=0,1$ as a function of the logarithm of magnetic field strength and $E_{\nu}=15$ MeV for $\rho/\rho_0=3$. See text for details.} 
\label{Fig9}
\end{center}
\end{figure}

In order to see the differences induced by the use of several parametrizations of the Skyrme force we plot in Fig.~\ref{Fig8} the spin response function versus energy transfer for the Skyrme SLy4 (solid line) and SLy7 (dotted line) models at $B=5 \times 10^{17}$ G in the RPA dipolar approximation, $\rho/\rho_0=3$ and $q=0.1\, fm^{-1}$. There is a shift in strength and energies of different sign for both parametrizations in the high and low energy modes excited due to the fact that effective masses and polarization in the plasma are slightly modified for both interaction \cite{angprc}. Let us notice that the energy shifts belong to the sub-MeV range, the same order of the induced resolution of the degeneracy of the $\omega_{\sigma}$ modes in the polarized system.
In Fig.~\ref{Fig9} we have plot the ratio of the neutrino mean free paths $R_l=\frac{\lambda_B-\lambda_{B=0}}{\lambda_{B=0}}$ versus the logarithm (base 10) of the magnetic field strength in the RPA monopolar ($l=0$) and dipolar ($l \le 1$) approximation for the Skyrme SLy4  and SLy7 parametrizations at a density $\rho/\rho_0=3$ and and incoming neutrino energy of $E_{\nu}=15$ MeV. Since this ratio stays positive in the $x$-axis the  neutrino opacity is smaller as magnetic field increases. We can see that for the Skyrme SLy4 parametrization at this density there is a negligible difference in the monopolar and dipolar case and the decrease in the opacity with respect to the $B=0$ case is  $\approx 4\%$. For the Skryme SLy7 parametrization we can see that the change in the dipolar case is smaller than in the monopolar case as magnetic field increases. At the maximum magnetic field strength considered at this density there is a $10\%$ decrease in opacity with respecto to the non-magnetized case in the dipolar response. 

\begin{figure}[hbtp]
\begin{center}
\includegraphics [angle=-90,scale=.75] {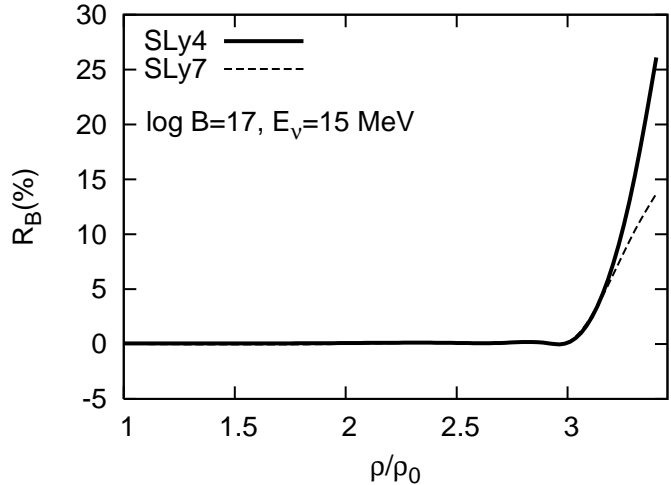}
\caption{Ratio of neutrino mean free paths as a function of density calculated with the SLy4 and SLy7 parametrizations for a value of magnetic field strength $B=10^{17}$ G and a neutrino energy $E_{\nu}=15$ MeV.} 
\label{Fig10}
\end{center}
\end{figure}

In order to check the density dependence of the relative changes induced by the inclusion of magnetic fields we plot in Fig.~\ref{Fig10} the ratio of neutrino mean free paths in the RPA dipolar approximation computed for a fixed value of magnetic field stregth $B=10^{17}$ G with respect to the $B=0$ G case, $R_B=\frac{\lambda_{B=10^{17}G} - \lambda_{B=0}}{\lambda_{B=0}}$ as a function of the density with the Skyrme SLy4 (solid line) and SLy7 (dashed line) parametrizations for a value of the neutrino incoming energy $E_{\nu}=15$ MeV. We can see that the SLy4 predicts a larger decrease in the neutrino opacity than for the SLy7 case as the density grows to approach a ferromagnetic transition. This behaviour is in agreement with the dramatic decrease of the neutrino mean free path as a phase transition arises in the system as obtained for the non-magnetized systems \cite{vida1}.

\section{Summary and conclusions}
\label{summary}

We have investigated in the context of the Landau Theory of normal Fermi Liquids, the effect of a strong magnetic field on the density (spin density) fluctuations in a two-component partially magnetized pure neutron system  within the framework of the non-relativistic Hartree-Fock approximation. To describe the nuclear interacting system we have used Skryme effective forces under the form of the SLy4 and SLy7 parametrizations. Due to the tiny value of the neutron magnetic moment we have considered magnetic field strengths up to the limiting value $B \approx 10^{18}$ G as allowed by the scalar virial theorem. The matter vector and vector-axial response functions, $S^{(S=0,1)}$ are obtained from the imaginary part of the Lindhard function, calculated in the RPA approximation using monopolar $(l=0)$ and dipolar $(l \le 1)$ expansion terms in the polarized quasiparticle interaction matrix element. The dipolar approximation is exact for the Skryme force and adds non-negligible contributions to the usually considered monopolar RPA structure functions, shifting the plasma collective modes to the lower (higher) energy range of the spectrum in the vector (vector-axial) functions with respect to the monopolar case. The collective modes, $\omega_{\sigma}$, depend on the net polarization so that when it is not vanishing the degeneracy of the modes in the non-magnetized case is resolved into pairs $\omega_+, \omega_-$. When the polarization is mild there is a sub-MeV energy mode separation increasing as density and magnetic field strength grows. As most of Skyrme models, near the predicted ferromagnetic transition a low energy mode arises, signaling the energetic instability towards the $\Delta=-1$ case.

In order to size the contribution to the results of the inclusion of a magnetic field we have also computed the ratios of variation of the collective modes with respect to the $B=0$ G case finding that the as density and magnetic field strength increases, the relative splitting grows. Using different Skyrme parametrizations in the NN interaction produces a sub-MeV shift in the collective mode location. Finally, to understand the response to a weak probe interacting via neutral currents we find that the magnetic field tends to decrease neutrino opacities giving a non-negligible $\approx 10\%$ correction or higher for a density of about three times saturation density at the maximum magnetic field strength $log_{10} B(G)=18$ and incoming neutrino energy $E_{\nu}=15$ MeV.
The inclusion of finite temperature magnetized matter in the presence of magnetic fields should be taken into account in astrophysical scenarios by considering beta-equilibrated matter so that polarization of protons would include corrections to the present result. This task should be addressed in future work. 

\vspace{2ex}

\noindent{\bf Acknowledgments}\\

We acknowledge discussions with J. Navarro and A. Polls. This work has been partially funded by the Spanish Ministry of Science and Education under project DGI-FIS2006-05319.


\end{document}